\documentclass[print,twocolumn,showpaces,aps,showpacs]{revtex4}
\usepackage{amsfonts}
\usepackage{amsmath}
\usepackage{amssymb}
\usepackage{graphicx}

\begin{document}

\title{Multisoliton Newton's cradles and supersolitons in regular and $%
\mathcal{PT}$-symmetric nonlinear couplers }
\author{Pengfei Li$^{1}$, Lu Li$^{1}$ }
\email{llz@sxu.edu.cn}
\author{Boris A. Malomed$^{2}$}
\affiliation{$^{1}$Institute of Theoretical Physics, Shanxi University, Taiyuan 030006,
China}
\affiliation{$^{2}$Department of Physical Electronics, School of Electrical Engineering,
Faculty of Engineering, Tel Aviv University, Tel Aviv 69978, Israel }

\begin{abstract}
We demonstrate the existence of stable collective excitation in the form of
``supersolitons" propagating through chains of solitons with alternating
signs (i.e., Newton's cradles built of solitons) in nonlinear optical
couplers, including the $\mathcal{PT}$-symmetric version thereof. In the
regular coupler, stable supersolitons are created in the cradles composed of
both symmetric solitons and asymmetric ones with alternating polarities.
Collisions between moving supersolitons are investigated too, by the means
of direct simulations in both the regular and $\mathcal{PT}$-symmetric
couplers.
\end{abstract}

\maketitle

\section{Introduction and the model}

Solitons offer a straightforward realization of the wave-particle dualism in
classical settings. Indeed, being built as self-trapped wave pulses, they
often behave as particles \cite{Scott}. Well-known applications of optical
solitons to data transmission and (potentially) all-optical data processing
make it necessary to design various handling devices, one of basic types of
which is the nonlinear directional coupler. Nonlinear twin-core fiber
couplers have been a subject of intensive studies since they were proposed
in Refs. \cite{Maier,Jensen}. Soliton dynamics in them has been analyzed in
detail, especially as concern the use of the couplers for the switching of
narrow pulses, both theoretically \cite%
{OL13/672,OL14/131,dyn/Romangoli,JOSAB1379,OL18/328,Switch/Uzunov,Manolo}
and experimentally \cite{APL1135,OL13/904}. Stationary modes supported by
the couplers have been studied too, revealing the existence of symmetric
\cite{PRA4455,PRA6278}, antisymmetric, and asymmetric solitons \cite%
{JOSAB1379,PRL2395,Nail4710,ASC}. Collisions between solitons of different
types were investigated in Ref. \cite{Scripta}. In addition to the symmetric
twin-core couplers, static and dynamical modes and their stability were
explored in asymmetric couplers \cite{Boris4084,Dave}, including those with
opposite signs of the group-velocity dispersion \cite{Dave2,Brenda}.

The analysis of the stability and interactions of solitons has been extended
to active dual-core systems, such as the model with gain and loss acting in
two coupled cores \cite{Winful,PRE4371,Chaos}. Recently, a $\mathcal{PT}$%
-symmetric extension of the concept of solitons in nonlinear couplers was
introduced, assuming, as it should be in diverse realizations of such
systems \cite%
{PT1,OL2632,PT3,PT4,PRA010103,OL2976,PT5,PT6,PT7,PT8,PRE046609,PT9,PT10,PT11,PT12,PT13,PT14,PT15,PT16}%
, the balance between the gain and loss in the two cores of the coupler \cite%
{DribenOL4323,PTcoupler2,PTcoupler3,PRA063837,PTcoupler4} (in the
above-mentioned earlier studied models, the loss acting in the passive core
was stronger than the gain in the active one, to secure the stability margin
for the resulting dissipative solitons \cite{Winful,PRE4371,Chaos}).

The quasi-particle properties of solitons suggest to employ them for
emulating various effects from classical mechanics in optical settings, with
possible applications to the design of compact nonlinear-optical circuitry.
One recently elaborated example is a possibility to use arrays of optical
solitons, in dissipative two-dimensional \cite{PRE012916} and conservative
one-dimensional \cite{Driben063808} setups alike, for building optical
counterparts of the Newton's cradle (NC), which are well known in mechanics
and molecular dynamics \cite{cradle1,cradle2,cradle3,cradle4}, and
\textquotedblleft supersolitons", i.e., self-supporting dislocations
propagating in chains of individual solitons. Previously, \textquotedblleft
supersolitons" were experimentally realized and theoretically studied in
chains of fluxons populating long periodically inhomogeneous Josephson
junctions \cite{Lyosha1,Lyosha2}, and predicted in binary Bose-Einstein
condensates (BECs), with attractive interactions in each component and
repulsion between them \cite{Supersoliton}. In this work, we aim to develop
a robust NC model, using soliton chains in nonlinear couplers, and construct
stable supersolitons as localized collective excitations in such chains.

The propagation of optical pulses in a nonlinear dual-core fiber coupler can
be described by two linearly coupled nonlinear Schr\"{o}dinger (NLS)
equations. This model was applied to the soliton switching in erbium-doped
nonlinear fiber couplers \cite{OL16/1653} and passively mode-locked lasers
\cite{OL17/1688}, and to the prediction of stable solitons in two-component
active systems \cite{Winful}. In this paper, we start with the coupled NLS
equations similar to those adopted in Ref. \cite{PRE4371}:%
\begin{align}
iU_{Z}+(1/2)U_{TT}+\left\vert U\right\vert ^{2}U+KV& =i\Gamma _{1}U,
\label{equ1a} \\
iV_{Z}+(1/2)V_{TT}+\left\vert V\right\vert ^{2}V+KU& =-i\Gamma _{2}V,
\label{equ1b}
\end{align}%
where $U(Z,T)=NA_{U}(z,t)/\sqrt{P_{0}}$\ and $V(Z,T)=NA_{V}(z,t)/\sqrt{P_{0}}
$\ are the normalized modal field amplitudes in the two cores, $Z=z/L_{D}$\
and $T=(t-z/v_{g})/T_{0}$\ are the normalized length and time in soliton
units, $A_{U,V}(z,t)$\ are slowly varying amplitudes of the electromagnetic
waves \cite{Agrawal}, $K=L_{D}C$\ is proportional to the linear coupling
coefficient $C$ of the dual-core fiber, while $\Gamma _{1}=gL_{D}/2$ and $%
\Gamma _{2}=\alpha L_{D}/2$\ \ represent the normalized linear gain and loss
in the active and passive cores. Further, $T_{0}$\ is the width of the
incident pulse, $v_{g}$\ is the group velocity, $L_{D}$, $L_{NL}$\ and $N$\
are, respectively, the dispersion length, nonlinearity length, and soliton
order, defined as
\begin{equation}
L_{D}=T_{0}^{2}/\left\vert \beta _{2}\right\vert ,L_{NL}=1/(\gamma P_{0}),\
N=\sqrt{L_{D}/L_{NL}},  \label{L}
\end{equation}%
where $\beta _{2}<0$\ and $\gamma $\ are the group-velocity dispersion (GVD)
and effective Kerr coefficient of the fiber, and, finally, $g$\ and $\alpha $%
\ are the effective gain, provided by dopants in the active core, and loss
in the passive one \cite{PRA44/7493}. The same equations with $g=\alpha $,
i.e., $\Gamma _{1}=\Gamma _{2}$, represent the model of the parity-time ($%
\mathcal{PT}$)-symmetric coupler, with mutually balanced gain and loss \cite%
{DribenOL4323}. The present model neglects higher-order effects, such as the
third-order dispersion and the shock and Raman terms.

The usual model of the symmetric coupler implies the that the propagation
length is much smaller than the dissipation length, hence $\Gamma
_{1}=\Gamma _{2}=0$ is assumed, and Eqs. (\ref{equ1a}) and (\ref{equ1b}) can
be rescaled ($U\equiv \sqrt{K}u$, $V\equiv \sqrt{K}v$, $T\equiv \tau /\sqrt{K%
}$, $\zeta \equiv KZ$) to a form devoid of any free parameter:%
\begin{align}
iu_{\zeta }+(1/2)u_{\tau \tau }+\left\vert u\right\vert ^{2}u+v& =0,
\label{u} \\
iv_{\zeta }+(1/2)v_{\tau \tau }+\left\vert v\right\vert ^{2}v+u& =0.
\label{v}
\end{align}

While Eqs. (\ref{equ1a}) and (\ref{equ1b}) are written in terms of the
temporal-domain propagation, which corresponds to the couplers in the form
of twin-core nonlinear optical fibers, the model may be realized in the
spatial domain as well, pertaining to a twin-core planar optical waveguide,
with temporal variable $\tau $ replaced by transverse coordinate $x$. The
same system, with $\zeta $ and $\tau $ replaced, respectively, by time and $%
x $, plays the role of Gross-Pitaevskii equations for the BEC trapped in a
dual-core prolate waveguide \cite{Arik}.

\begin{figure}[tbp]
\centering\vspace{-0.0cm} \includegraphics[width=8cm]{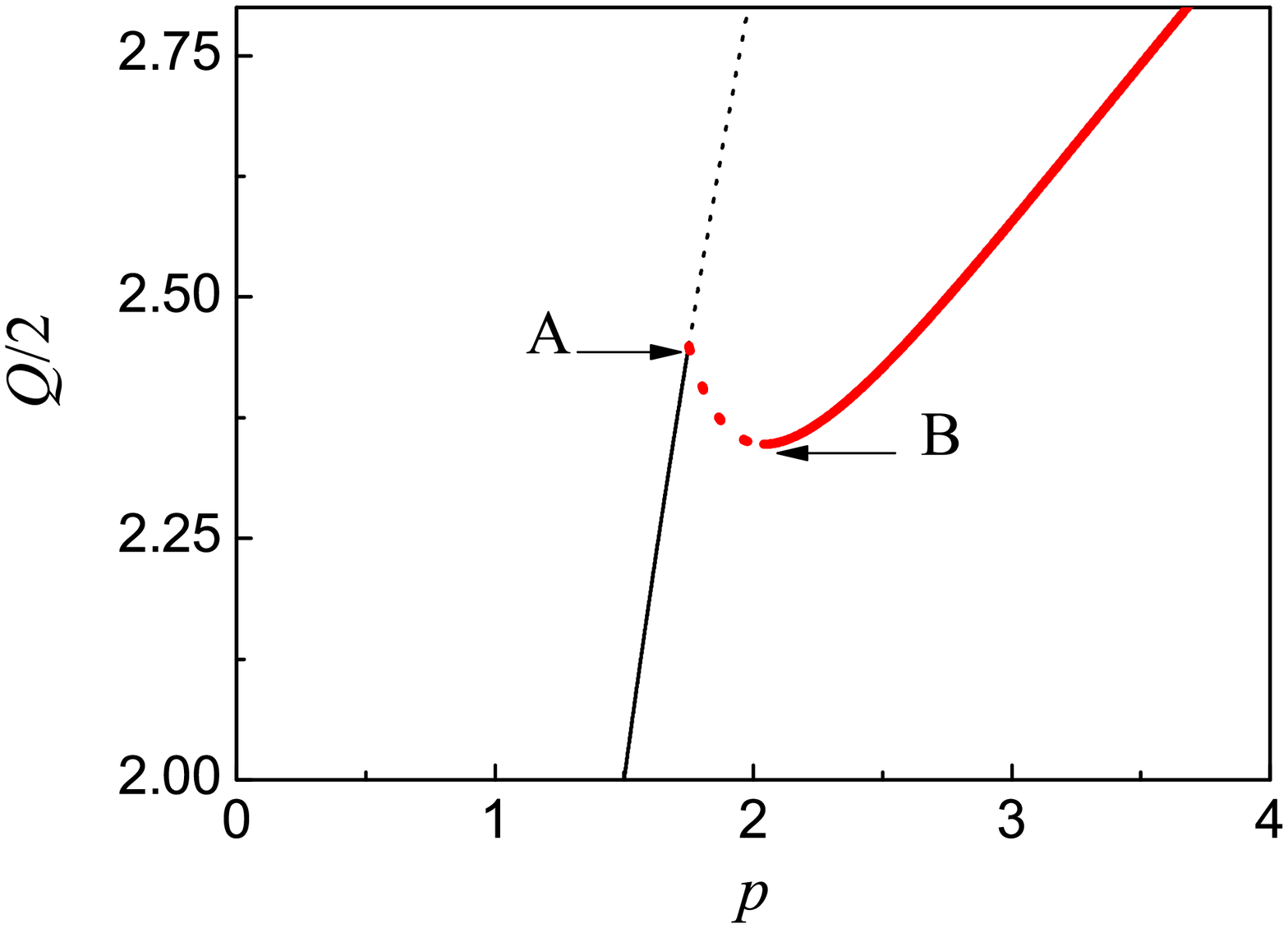} \vspace{%
-0.0cm}
\caption{(Color online) The bifurcation diagram for solitons in the
symmetric coupler (without gain and loss), in terms of the dependence
between the total energy ($Q$) and propagation constant ($p$). The diagram
was produced by means of the variational approximation in Ref. \protect\cite%
{Boris4084}. The solid and dot curves correspond to stable and unstable
solutions, respectively, while the light black and thick red curves
represent, severally, symmetric and asymmetric solitons. $Q_{\text{s}}$
(point A) is the point of the subcritical symmetry-breaking bifurcation,
while $Q_{\text{a}}$ (point B) indicates the related tangent bifurcation
which gives rise to the pair of stable and unstable asymmetric solitons.}
\label{fig1}
\end{figure}

It is well known that the system of equations (\ref{u}), (\ref{v}) gives
rise to symmetric and asymmetric soliton solutions \cite%
{OL13/672,OL14/131,dyn/Romangoli,JOSAB1379,OL18/328,Switch/Uzunov,Manolo,APL1135,OL13/904,PRA4455,PRA6278,PRL2395,Nail4710,ASC,Scripta,Boris4084}%
, whose stability depends on the total energy (norm),
\begin{equation}
Q=\int_{-\infty}^{+\infty}\left( \left\vert u\right\vert ^{2}+\left\vert
v\right\vert ^{2}\right) d\tau\equiv Q_{u}+Q_{v},  \label{Q}
\end{equation}
or, in other words, on the corresponding propagation constant, $p$, as shown
by means of the $Q(p)$ dependence in Fig. \ref{fig1} (it actually reproduces
Fig. 2 from Ref. \cite{Boris4084}). The transition from symmetric to
asymmetric solitons, with the increase of $Q$, occurs via a subcritical
symmetry-breaking bifurcation, which features a narrow region of bistability.

The rest of the paper is organized as follows. In Section II, collective
excitations in the form of supersolitons in chains of symmetric and
asymmetric solitary pulses are studied in the model of the ordinary
symmetric coupler. The analysis of the $\mathcal{PT}$-symmetric
generalization of the model is reported in Section III. Conclusions are
presented in Section IV.

\section{Supersolitons in soliton chains in the nonlinear coupler}

Symmetric and antisymmetric solitons are represented by obvious exact
solutions to Eqs. (\ref{u}) and (\ref{v}), which reduce to the classical NLS
(nonlinear-Schr\"{o}dinger) solitons \cite{JOSAB1379}:%
\begin{equation}
u=\pm v=\sqrt{2\beta}\mathrm{sech}\left( \sqrt{2\beta}\tau\right) \exp\left[
i\left( \beta\pm1\right) \zeta\right] ,  \label{exact_solu_1}
\end{equation}
where $\beta\pm1$ are propagation constants of the symmetric and
antisymmetric solutions, respectively. For the symmetric solution, $\beta+1$
is the propagation constant which is denoted as $p$ in Fig. \ref{fig1}. It
is well known that the symmetric solution is stable at
\begin{equation}
\beta\leq\beta_{\max}=2/3,  \label{2/3}
\end{equation}
and unstable at $\beta>\beta_{\max}$ \cite{PRA4455}. The total energy of
solitons (\ref{exact_solu_1}) is $Q=4\sqrt{2\beta}$, hence the corresponding
critical value is%
\begin{equation}
Q_{\text{s}}^{\mathrm{(exact)}}=8/\sqrt{3}\approx4.62,  \label{16/3}
\end{equation}
Note that Fig. \ref{fig1} presents a counterpart of this result predicted by
the variational approximation, $Q_{\text{s}}^{\mathrm{(var)}}=2\sqrt{6}%
\approx\allowbreak4.90$ \cite{Boris4084}. Because, as mentioned above, the
symmetry-breaking bifurcation is subcritical, pairs of stable and unstable
asymmetric solitons appear, via the tangent bifurcation at $Q>Q_{\text{a}}$,
see Fig. \ref{fig1}. Unlike the exact critical point (\ref{16/3}), the value
of $Q_{\text{a}}$ is known in an approximate form, produced by the
variational method: $Q_{\text{a}}^{\mathrm{(var)}}=3\cdot6^{1/4}\approx%
\allowbreak4.70$ \cite{Boris4084}, while its numerically generated
counterpart can be taken from Fig. 11 of Ref. \cite{ASC} which is somewhat
smaller,
\begin{equation}
Q_{\text{a}}^{\mathrm{(num)}}\approx4.58.  \label{Qa}
\end{equation}
A narrow region of the \textit{bistability} of the asymmetric and symmetric
solitons is $Q_{\text{a}}^{\mathrm{(num)}}<Q<Q_{\text{s}}^{\mathrm{(exact)}}$%
. For the antisymmetric solitons, the effective stability area is much
smaller than for the symmetric ones \cite{Nail4710}, and, strictly speaking,
all the antisymmetric solitons are subject to weak instability \cite%
{PRA063837}, therefore we do not consider antisymmetric solutions below.

Using stable individual symmetric solitons, whose energy does not exceed the
threshold value (\ref{16/3}), one can construct a soliton chain with
alternating signs of adjacent solitons:%
\begin{equation}
\left( u,v\right) ,\left( -u,-v\right) ,\left( u,v\right) ,\cdots ,\left(
-1\right) ^{n-1}\left( u,v\right) .  \label{uni_array}
\end{equation}
The opposite signs are necessary to guarantee repulsion between neighboring
solitons, otherwise the chains will be obviously unstable. Experimentally,
the alternating signs of temporal solitons in the dual-core fiber can be
provided by a modulator combined with the laser source generating the pulses.

Using these chains, one can initiate the NC dynamics by \textit{kicking} one
soliton, i.e., multiplying its both components by $\exp (ik\tau )$ with kick
strength $k$. The Galilean invariance of Eqs. (\ref{u}) and (\ref{v})
implies that $\mathrm{sech}(\sqrt{2\beta }\tau )$ in solution (\ref%
{exact_solu_1}) is then replaced by $\mathrm{sech}[\sqrt{2\beta }(\tau
-k\zeta )]$, i.e., the kicked soliton is \textit{boosted} at rate $k$. Thus
the configuration represented by Eq. (\ref{uni_array}) gives rise to the NC
built as the chains of optical solitons, where the kicked one plays the role
of the impacting ball in the mechanical realization of the NC.

\begin{figure}[tbp]
\centering\vspace{-0.0cm} \includegraphics[width=9.0cm]{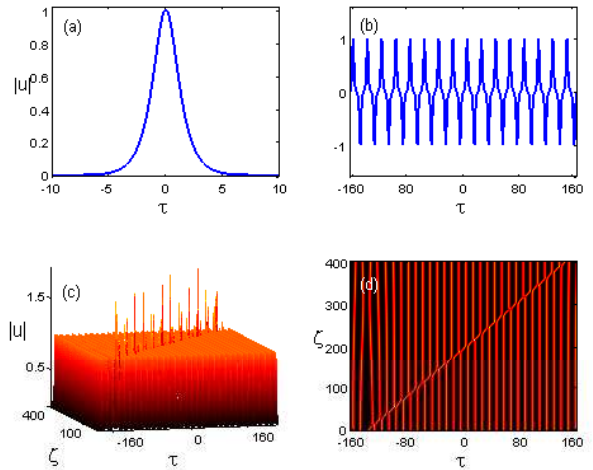} \vspace{%
-0.3cm}
\caption{(Color online) The dynamics of the Newton's cradle in the chain of
identical symmetric solitons hosted by the coupler. (a) An individual
symmetric solution with total energy $Q=4$. (b) The chain built of such
solitons with initial separation between adjacent solitons{\protect\LARGE \ }%
$\bigtriangleup \protect\tau =10$ and $n=32$. (c) The evolution of the chain
excited by boosting the third soliton with strength $k_{3}=0.5$. (d) The top
view corresponding to (c). In all panels, only the $u$ component of the
mutually symmetric fields $\left( u,v\right) $ is displayed.}
\label{fig2}
\end{figure}

The soliton chain with free edges will spontaneously expand because of the
repulsion between solitons, therefore the simulation were run in the system
with periodic boundary conditions, which may physically correspond to
optical solitons in a fiber loop \cite{loop,loop2}, or to the BEC in a
toroidal trap \cite{torus,torus2,torus3}. Figure \ref{fig2} presents results
of simulations of Eqs. (\ref{u}) and (\ref{v}) for the evolution of the
chain of symmetric solitons, excited by boosting the third soliton. The
figure clearly demonstrates that the momentum, which was originally imparted
to the third soliton, is transferred along the chain by particle-like
elastic collisions between the solitons. The collective excitation mode
propagating along the soliton chains may be identified as a
\textquotedblleft supersoliton", i.e., a self-supporting localized
perturbation moving along the chains of \textquotedblleft
primary\textquotedblright\ solitons \cite{Supersoliton}.

\begin{figure}[tbp]
\centering\vspace{-0.0cm} \includegraphics[width=9.0cm]{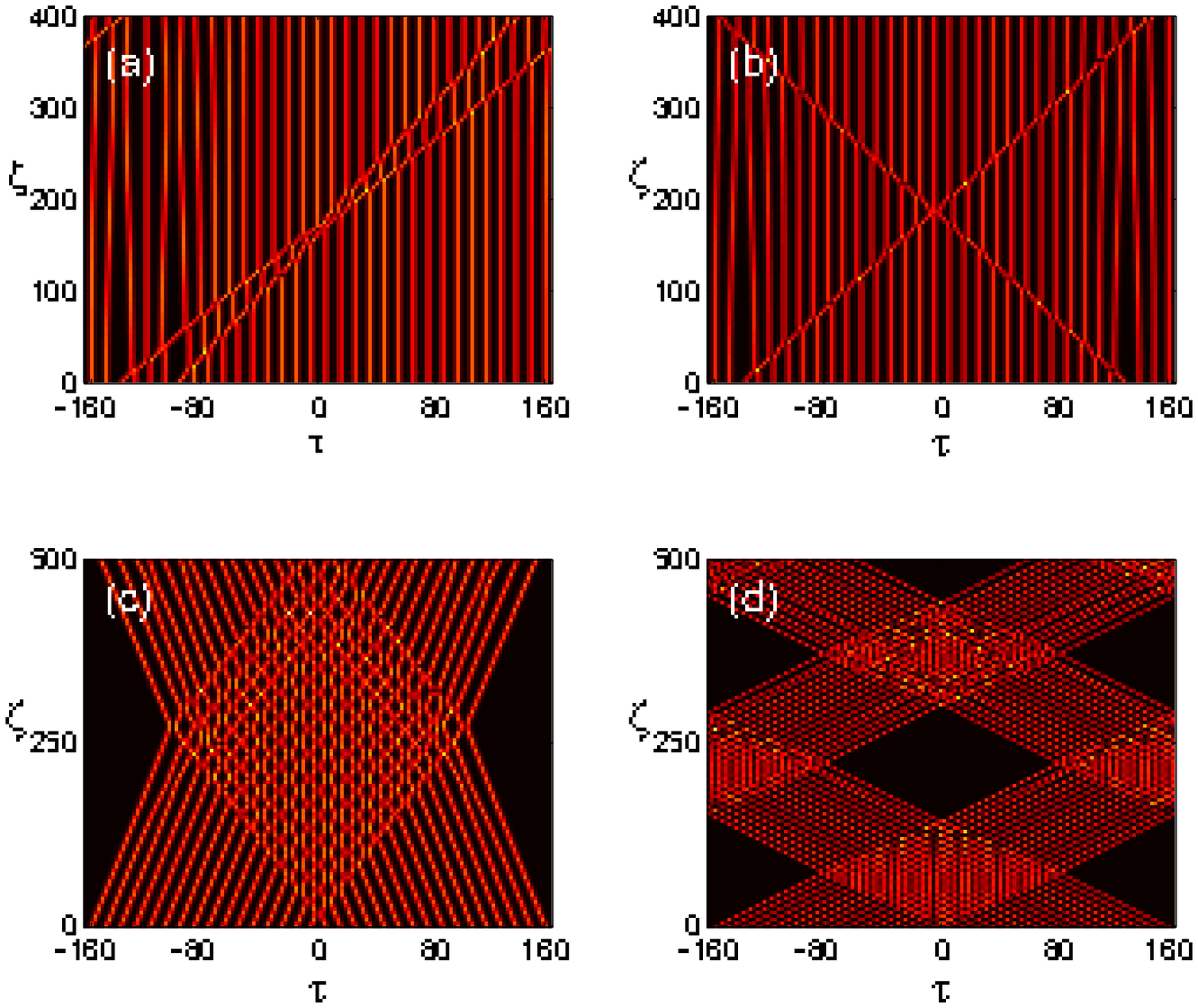} \vspace{%
-0.3cm}
\caption{(Color online) Collisions between supersolitons created in the same
chain as in Fig. 2. (a) An overtaking collision of supersolitons excited by
kicks with $k_{3}=0.6$ and $k_{7}=0.4.$ (b) A head-on collision excited by $%
k_{3}=-k_{29}=0.5.$ (c,d) Cascades of multiple collisions when the kick is
applied as follows: (c) $k_{1}=\cdots =k_{16}=0.2$ and $k_{17}=\cdots
=k_{32}=-0.2$; (d) $k_{1}=\cdots =k_{16}=1$ and $k_{17}=\cdots =k_{32}=-1$.
Other parameters are the same as in Fig. \protect\ref{fig2}.}
\label{fig3}
\end{figure}

Collisions between two supersolitons are further illustrated in the Figs. %
\ref{fig3}(a) and (b), which show elastic overtaking and head-on collisions
in the chain built of symmetric solitons. Multi-supersoliton collisions may
be also initiated by kicking all solitons in the chains, as shown in Figs. %
\ref{fig3}(c) and (d). The character of the excitation of the chain in the
configuration displayed in Fig. \ref{fig3}(c) makes the number of collisions
for each soliton a function of its original position, the solitons located
closer to central position colliding a larger number of times. The
simulations performed with a larger value of the kick strength demonstrate
that the left and right parts of the chain perform collectively recurrent
elastic collisions, as shown in Fig. \ref{fig3}(d).

\begin{figure}[tbp]
\centering\vspace{-0.0cm} \includegraphics[width=9.0cm]{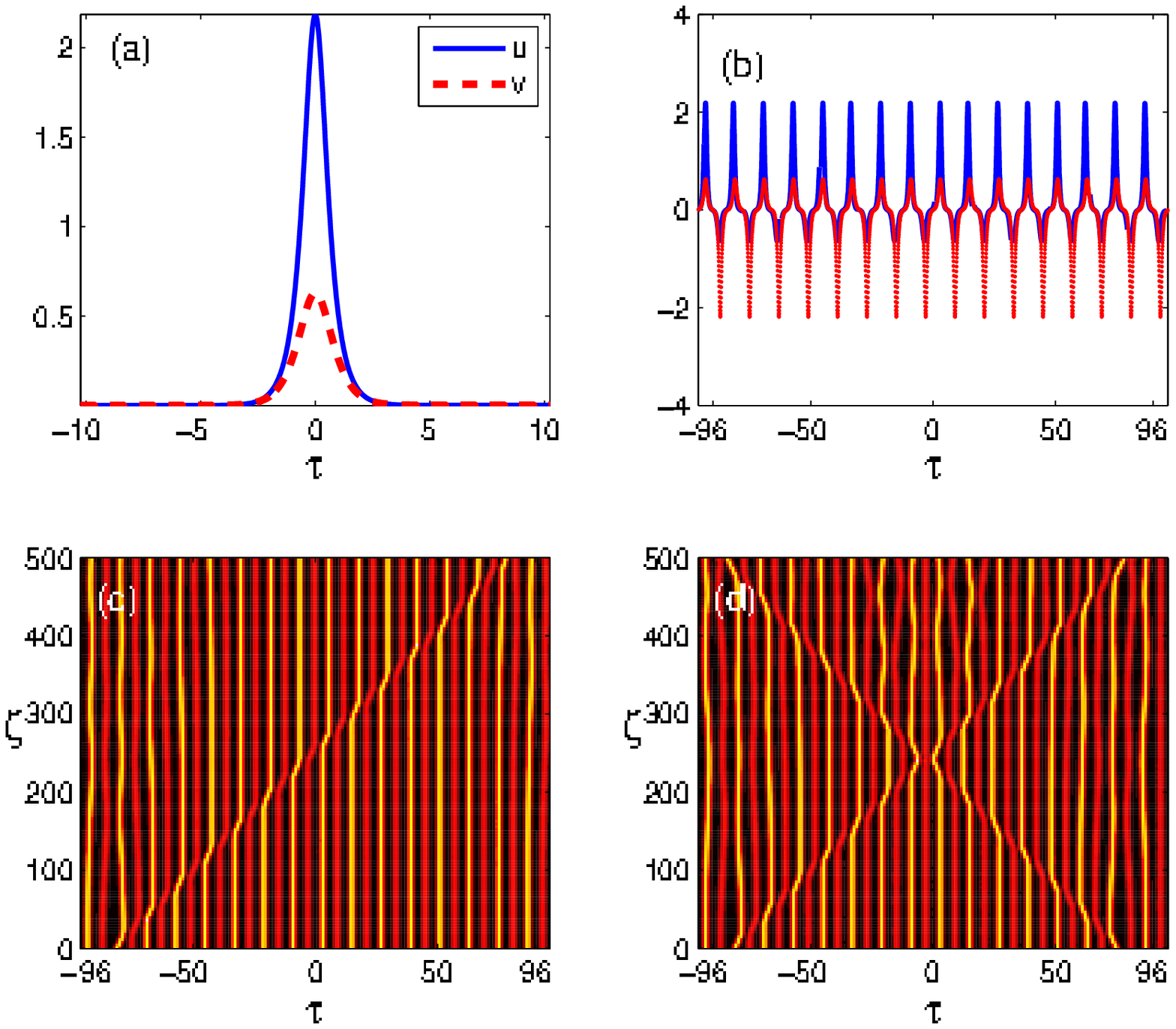} \vspace{%
-0.3cm}
\caption{(Color online) (a) Profiles of the two field components $u$ (the
solid blue line) and $v$ (dashed red line) of the stable asymmetric soliton
with total energy $Q=5$. (b) Chains of asymmetric solitons with alternating
polarities and opposite signs of adjacent solitons in the $u$-core (solid
blue line) and $v$-core (dot red line), with the initial separation between
adjacent solitons $\bigtriangleup \protect\tau =6$ and $n=32$. (c) The
propagation of a supersoliton, initiated by the application of the kick with
$k_{3}=0.15$ to the third soliton in array (\protect\ref{uni_array}). Here
components $|u|$ and $|v|$ are shown by yellow and red, respectively. (d)
The head-on collision between supersolitons created in the same chain as in
(c), by the application of kicks $k_{3}=-k_{29}=0.15$.}
\label{fig4}
\end{figure}

As said above, the symmetric solitons are unstable at $Q>Q_{\text{s}}$ [see
Eq. (\ref{16/3})], therefore in this case it is relevant to build chains of
asymmetric solitons. Because exact solutions for asymmetric modes are not
available, they can be created, using the initial guess provided by the
variational approximation, which was developed in Ref. \cite{Boris4084} on
the basis of a simple \textit{ansatz},%
\begin{equation}
u=A\cos \left( \theta \right) \mathrm{sech}\left( \tau /a\right) ,v=A\sin
\left( \theta \right) \mathrm{sech}\left( \tau /a\right) ,  \label{asantz_u}
\end{equation}%
with amplitude $A$, width $a$, and angle $\theta $ which determines the
asymmetry of the energy splitting between the cores, $Q_{v}/Q_{u}=\tan
^{2}\theta $. Starting from this ansatz (with parameters $A$, $a$ and $%
\theta $ taken from Ref. \cite{Boris4084}), numerically exact asymmetric
solutions were constructed by means of the Newton's method, as shown in Fig. %
\ref{fig4}(a,b). In panel (a) of the figure, ratios of the amplitudes and
energies in the two components of the asymmetric soliton are $%
A_{u}/A_{v}=3.48$ and $Q_{v}/Q_{u}=0.116$, respectively. The total energy of
the asymmetric soliton, $Q=5$, exceeds the minimum value (\ref{Qa}),
therefore it is stable. As mirror images of stable asymmetric solitons with $%
A_{u}/A_{v}>1$, there exist solitons with the opposite \textit{polarity},
i.e., $A_{u}/A_{v}<1$.

\begin{figure}[tbp]
\centering\vspace{-0.0cm} \includegraphics[width=9.0cm]{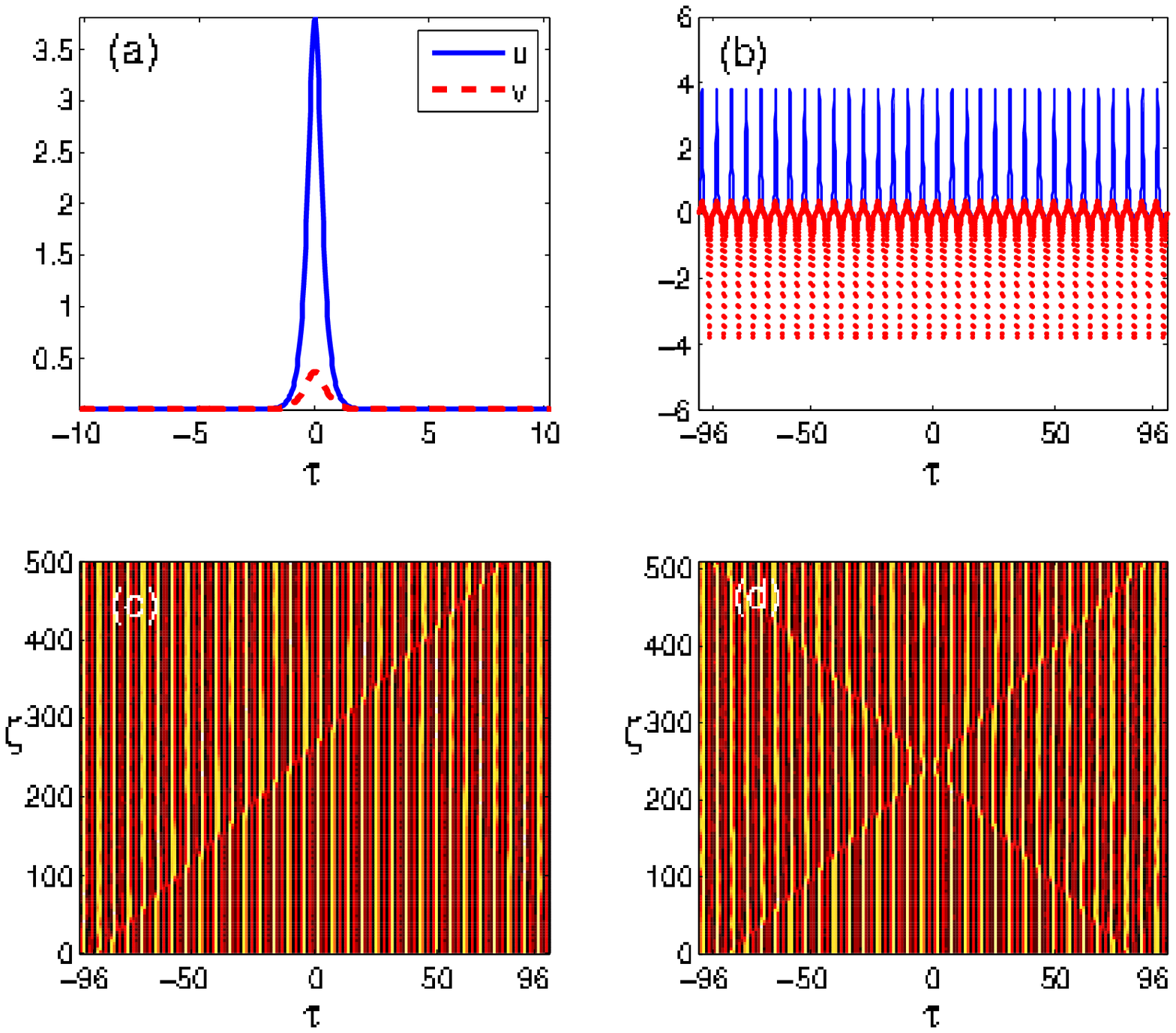} \vspace{%
-0.3cm}
\caption{(Color online) The same as in Fig. \protect\ref{fig4}, but for a
chain built of strongly asymmetric solitons, each with total energy $Q=7.74$%
. The separation between adjacent solitons and their number are $\Delta
\protect\tau =3$ and $n=64$. The initial kicks are $k_{3}=0.1$ in (c), and $%
k_{5}=-k_{59}=0.1$ in (d). Other parameters are the same as in Fig. \protect
\ref{fig4}.}
\label{fig5}
\end{figure}

Thus, a chain of asymmetric solitons with alternating polarities and phase
shifts $\pi $ between adjacent ones [cf. Eq. (\ref{uni_array})] can be build
in the form of%
\begin{equation}
\left( u,v\right) ,\left( -v,-u\right) ,\left( u,v\right) ,\left(
-v,-u\right) ,\cdots ,  \label{asy_array}
\end{equation}%
as shown in Fig. \ref{fig4}(b). Then, similar to the case of the chain
composed of symmetric solitons, the NC dynamics can be initiated in this
chain by kicking both components of a selected soliton see Fig. \ref{fig4}%
(c). A chain can also be composed of asymmetric solitons with identical
polarities, but that case seems less interesting. In this context, it is
relevant to mention that collision between a stable asymmetric soliton and
its counterpart with the opposite polarity was first studied, by means of
direct simulations, in Ref. \cite{Scripta}.

Figure. \ref{fig4}(c) demonstrate that collective supersolitons can be
created in the NC of the present type, if the strength of the initial kick
is selected appropriately. An elastic head-on collisions between two
supersolitons is displayed in Fig. \ref{fig4}(d).

\begin{figure}[tbp]
\centering\vspace{-0.0cm} \includegraphics[width=9.0cm]{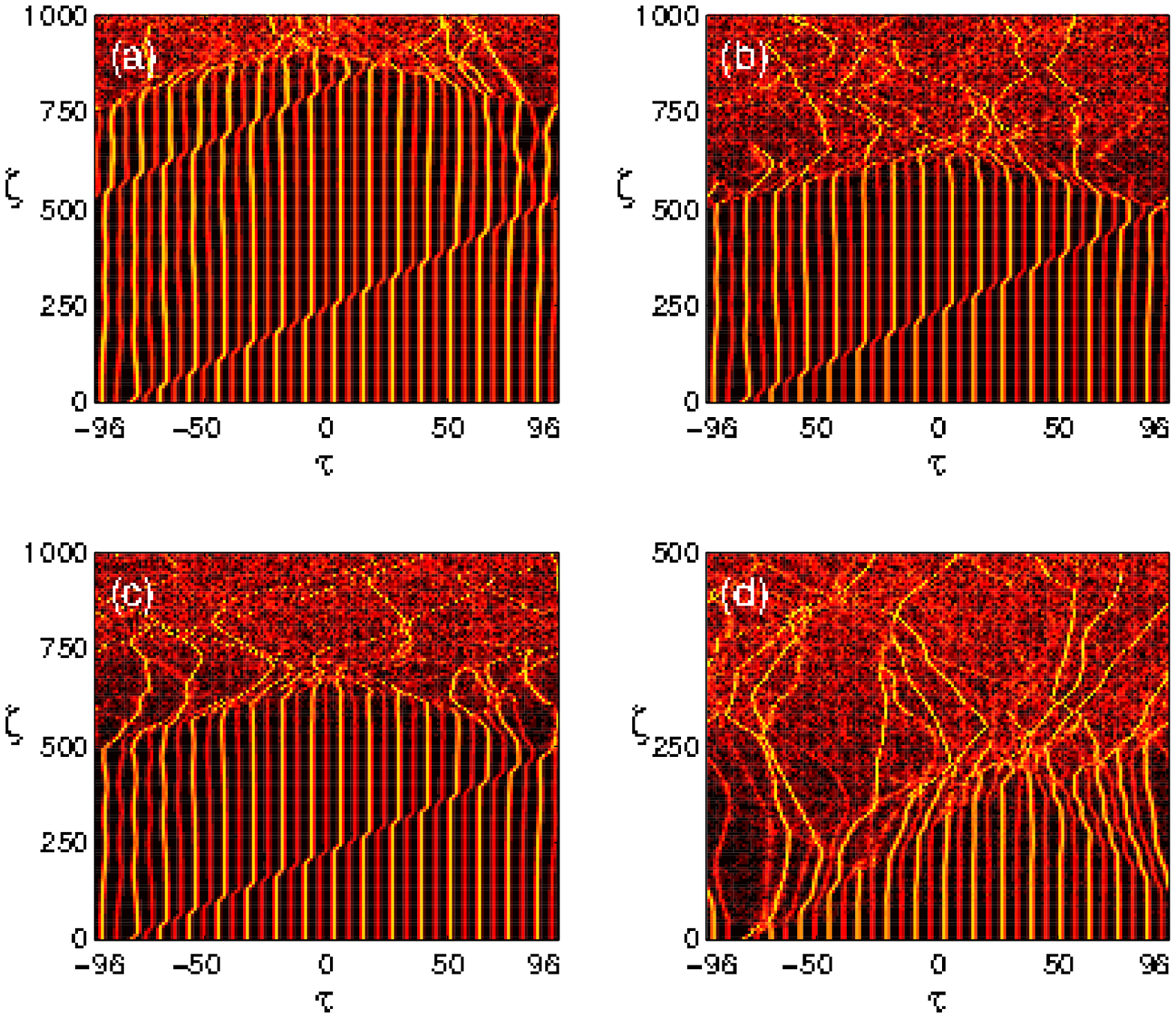} \vspace{%
-0.3cm}
\caption{(Color online) The evolutions of the chain built of asymmetric
solitons with the individual energy $Q=5$, for the different strengths of
the initial kick: (a) $k_{3}=0.16$, (b) $k_{3}=0.17$, (c) $k_{3}=0.18$ and
(d) $k_{3}=0.5$. The other parameters are the same as in Fig. \protect\ref%
{fig4}.}
\label{fig6}
\end{figure}

\begin{figure}[tbp]
\centering\vspace{-0.0cm} \includegraphics[width=9.0cm]{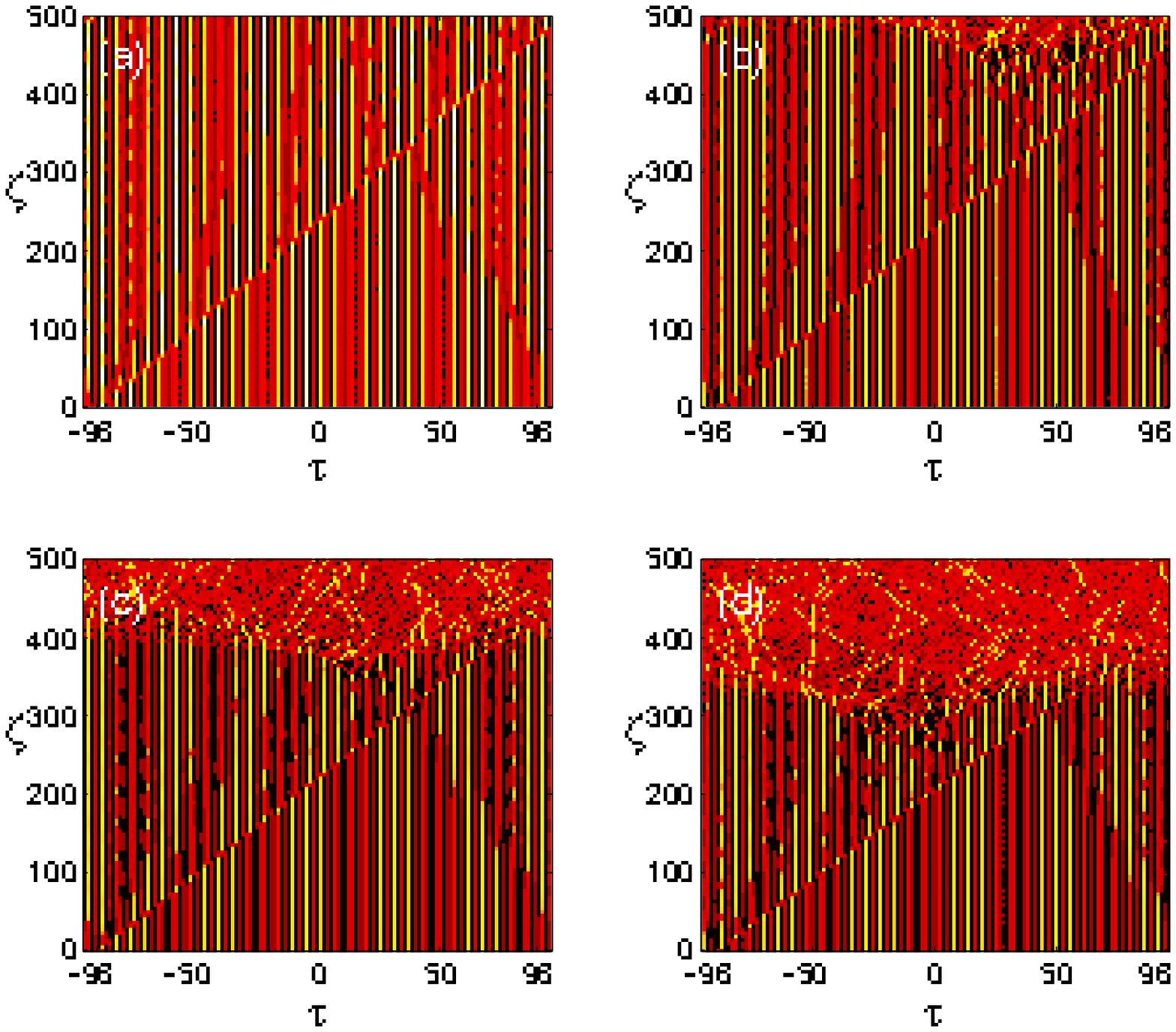} \vspace{%
-0.3cm}
\caption{(Color online) The same as in Fig. \protect\ref{fig6}, but for $%
Q=7.74$: (a) $k_{3}=0.13$, (b) $k_{3}=0.14$, (c) $k_{3}=0.15$ and (d) $%
k_{3}=0.17$. The other parameters are the same as in Fig. \protect\ref{fig5}%
. }
\label{fig7}
\end{figure}

The NC and supersolitons propagating in it can be built too in a chain of
strongly asymmetric solitons, as shown in Fig. \ref{fig5}, where the total
energy of the individual soliton is $Q=7.74$, with the ratios of the
amplitudes and energies of the two components being $A_{u}/A_{v}=10.5$ and $%
Q_{v}/Q_{u}=0.0134$, see Fig. \ref{fig5}(a). For an appropriately chosen
initial kick, the NC dynamics in the present chain is similar to that
observed in Fig. \ref{fig4}. In particular, the collision between
supersolitons, which is shown in \ref{fig5}(d), is again elastic.

The size of the initial kick is important for the excitation of
supersolitons in the chains of asymmetric solitons with alternating
polarities. Figures \ref{fig6} and \ref{fig7} display the evolution of the
chain with energies of individual solitons $Q=5$ and $Q=7.74$, for different
kicks. It is observed that the chains are destabilized by increasing the
strength of the kick. This is different from the chains of symmetric
solitons, which remain robust\ even under the action of strong kicks.

\section{The excitation of supersoliton in $\mathcal{PT}$-symmetric Coupler}

As shown in detail in Refs. \cite%
{DribenOL4323,PTcoupler2,PTcoupler3,PRA063837,PTcoupler4}, a $\mathcal{PT}$%
-symmetric coupler can be constructed, on the basis of the usual one
considered above, by adding linear-gain and loss terms with equal
coefficients ($\Gamma _{0}$) to Eqs. (\ref{u}) and (\ref{v}), respectively
[which corresponds to $\Gamma _{1}=\Gamma _{2}=\Gamma _{0}$ in Eqs. (\ref%
{equ1a}) and (\ref{equ1b})]:
\begin{align}
iu_{\zeta }+(1/2)u_{\tau \tau }+\left\vert u\right\vert ^{2}u+v& =i\Gamma
_{0}u,  \label{pt_u} \\
iv_{\zeta }+(1/2)v_{\tau \tau }+\left\vert v\right\vert ^{2}v+u& =-i\Gamma
_{0}v.  \label{pt_v}
\end{align}%
Further, the substitution of
\begin{equation}
\Phi (\tau ,\zeta )\equiv v(\tau ,\zeta )=\left( i\Gamma _{0}\pm \sqrt{%
1-\Gamma _{0}^{2}}\right) u(\tau ,\zeta ),  \label{Phi}
\end{equation}%
transforms Eqs. (\ref{u}) and (\ref{v}) into a single\ equation for $\Phi $:%
\begin{equation}
i\Phi _{\zeta }+(1/2)\Phi _{\tau \tau }+\left\vert \Phi \right\vert ^{2}\Phi
\pm \sqrt{1-\Gamma _{0}^{2}}\Phi =0,  \label{FS_NLS}
\end{equation}%
provided that $\Gamma _{0}\leq 1$, hence the $\mathcal{PT}$-symmetric and
antisymmetric solitons, which correspond, respectively, to the upper and
lower signs $\pm $ in Eqs. (\ref{u}) and (\ref{v}), are found as
\begin{equation}
\Phi =\sqrt{2\beta }\mathrm{sech}\left( \sqrt{2\beta }\tau \right) \exp %
\left[ i\left( \beta \pm \sqrt{1-\Gamma _{0}^{2}}\right) \zeta \right] .
\label{Sol_FS_NLS}
\end{equation}%
For the $\mathcal{PT}$-symmetric solitons, the boundary of the stability
region is
\begin{equation}
\beta \leq \beta _{\max }=2\sqrt{1-\Gamma _{0}^{2}}/3  \label{PT-stab}
\end{equation}%
\cite{DribenOL4323,PRA063837}, while asymmetric solitons do not exist in
this system. $\mathcal{PT}$-antisymmetric solitons [recall they correspond
to the lower sign $\pm $ in Eq. (\ref{Phi})] are, strictly speaking, always
unstable \cite{PRA063837}, but, in some domain [which is essentially smaller
than the stability area (\ref{PT-stab}) of the symmetric solitons], the
antisymmetric solitons seem practically stable in direct simulations, as the
underlying instability is weak \cite{DribenOL4323}.

\begin{figure}[tbp]
\centering\vspace{-0.0cm} \includegraphics[width=9.0cm]{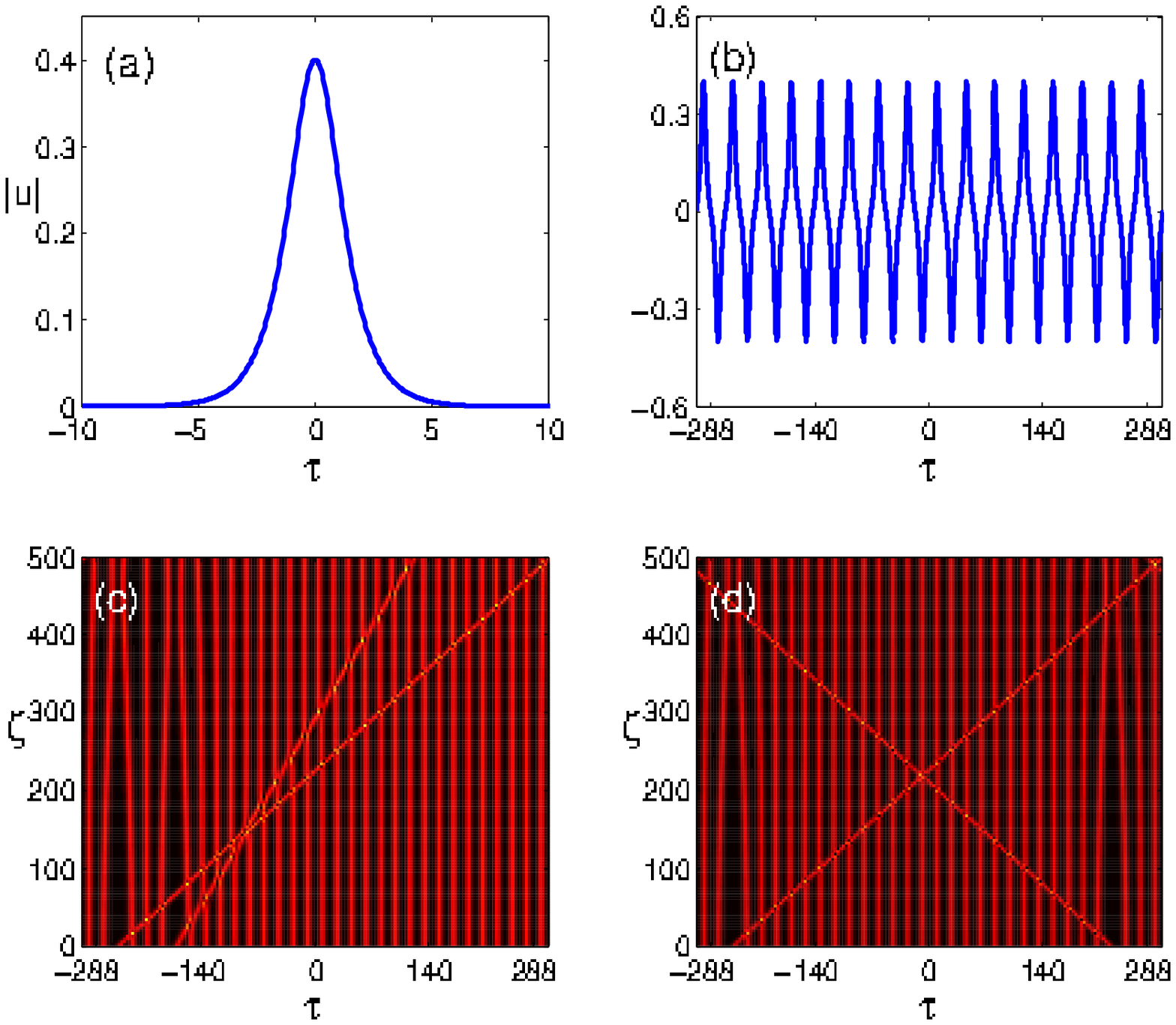} \vspace{%
-0.3cm}
\caption{(Color online) The supersoliton in the chain of symmetric solitons
housed by the $\mathcal{PT}$-symmetric coupler with the gain-loss
coefficient $\Gamma _{0}=0.5$. (a) The profile of an individual stable
symmetric soliton with total energy $Q=1.6$. (b) A chain composed of $32$
such solitons with initial separation between adjacent solitons%
{\protect\LARGE \ }$\bigtriangleup \protect\tau =18$. (c) An elastic
overtaking collision between the supersolitons excited by kicks $k_{3}=1$
and $k_{7}=0.5.$ (d) A head-on collision of the supersolitons excited by $%
k_{3}=-k_{29}=1$.}
\label{fig8}
\end{figure}

One can construct a $\mathcal{PT}$-symmetric soliton chains, and launch the
NC dynamics in it, following the same pattern as elaborated above. The
corresponding numerical results are displayed in Fig. \ref{fig8} (note that
the possibility of elastic collisions between individual stable $\mathcal{PT}
$-symmetric solitons was demonstrated in Ref. \cite{DribenOL4323}).

Figure \ref{fig8} demonstrates that the propagation of supersolitons in the $%
\mathcal{PT}$-symmetric coupler is qualitatively similar to that in the
usual nonlinear coupler, because collisions between individual solitons are
elastic. A difference from the soliton NC in the ordinary coupler is that
the application of the kick to a particular soliton creates a conspicuous
perturbation of the NC around that site, which then gradually relaxes. The
simulations also reveal the possibility of the propagation of supersolitons
in a chain composed of $\mathcal{PT}$-antisymmetric solitons (not shown
here), a natural difference being that the corresponding stability domain is
smaller than that in the chain of symmetric solitons.

\section{Conclusions and discussions}

In this work, we have demonstrated that the models of the usual and $%
\mathcal{PT}$-symmetric nonlinear couplers admit the realization of the NC
(Newton's cradle), in the form of a stable chain of optical solitons with
alternating signs, and the creation of stable \textquotedblleft
supersolitons\textquotedblright\ in the NC, in the form of self-supporting
localized collective excitations in the chains, propagating through
consecutive elastic collisions between individual solitons. In the usual
coupler, these results were demonstrated for the chains composed of both
symmetric solitons and asymmetric ones, with alternating polarities. The
elastic collisions between supersolitons were also found in usual and $%
\mathcal{PT}$-symmetric nonlinear couplers.

To estimate the predicted effects in physical units, we use real fiber
parameters to evaluate the corresponding propagation distances, pulse
widths, amplitudes, separations between adjacent solitons, the coupling
coefficient in the usual nonlinear couplers, built of cores with strong GVD
and Kerr nonlinearity (otherwise, the necessary propagation distance will be
too large), and the gain/loss constant in the $PT$-symmetric counterpart.
For this purpose, the GVD and Kerr coefficients of the fiber coupler are
taken as $\beta _{2}=-9.95\times 10^{2}$\ ps$^{2}/$km and $\gamma =10$W$%
^{-1} $km$^{-1}$ (such large values are available, e.g., in microstructured
fibers \cite{structured-fibers}), while the soliton order and dimensionless
coupling constant are set to be $N=K=1$. Thus, for a given initial power $%
P_{0}$, which should be high enough to secure a relatively short
nonlinearity distance (for example, $P_{\emph{0}}=10$ W), the
propagation-distance and temporal scales, $L_{NL}$, $L_{D}$\ and $T_{0}$,
can be identified [see Eq. (\ref{L})], and then the pulse amplitudes ($A_{U}$%
, $A_{V}$), widths ($W_{U}$, $W_{V}$), propagation distance $z$, coupling
coefficient $C$, and gain $g$, along with loss $\alpha $, may be retrieved
in physical unit. In this case, normalized propagation distance $\zeta =500$
in the figures corresponds to the fiber length $\sim 5$\ km, with the
coupling coefficient $\sim 0.1$\ m$^{-1}$. The the corresponding pulse
widths ($W_{U}$, $W_{V}$) in Figs. \ref{fig2} and \ref{fig3}, Figs. \ref%
{fig4} and \ref{fig6}, Figs. \ref{fig5} and \ref{fig7}, and Fig. \ref{fig8}
are, respectively, ($2.86$, $2.86$) ps, ($1.35$, $1.83$) ps, ($0.75$, $1.10$%
) ps, and ($7.1$, $7.1$) ps, while the temporal separations between adjacent
solitons are $31.5$\ ps, $19.0$\ ps, $9.5$\ ps, and $56.8$\ ps. For the $%
\mathcal{PT}$-symmetric coupler, the gain and loss coefficients are $0.43$\
dB/km. The propagation distance can be further reduced by increasing the
input power or using highly-nonlinear fibers \cite{OE11/3568}.

A challenging possibility is to generalize the analysis for chains and
lattices of regular \cite{Nir} and $\mathcal{PT}$-symmetric \cite{Gena}
spatiotemporal solitons in two-dimensional couplers with the cubic-quintic
nonlinearity, which support individual stable solitons of such types.\emph{\
}

\section{Acknowledgement}

This research was supported by the National Natural Science Foundation of
China, through Grant No.61078079, and by the Shanxi Scholarship Council of
China, through Grant No.2011-010.

\end{document}